\documentstyle[11pt,amssymb]{article}

\newcommand{\be}{\begin{equation}}
\newcommand{\ee}{\end{equation}}
\newcommand{\ba}{\begin{array}}
\newcommand{\ea}{\end{array}}
\newcommand{\bea}{\begin{eqnarray}}
\newcommand{\eea}{\end{eqnarray}}

\def\hbar{h\!\!\!/}

\begin{document}

\begin{titlepage}

\today\hfill PAR-LPTHE 95/49
\begin{flushright} hep-th/yynnmmm \end{flushright}
\vskip 4cm
\begin{center}
{\bf BILAYERS IN FOUR DIMENSIONS AND SUPERSYMMETRY}
\end{center}
\vskip .5cm
\centerline{Romain ATTAL
%    \footnote[0]{Member of `Centre National de la Recherche Scientifique'.}
     \footnote[1]{E-mail: attal@lpthe.jussieu.fr.}
            }
\vskip 5mm
\centerline{{\em Laboratoire de Physique Th\'eorique et Hautes Energies,}
     \footnote[2]{LPTHE tour 16\,/\,1$^{er}\!$ \'etage,
          Universit\'e P. et M. Curie, BP 126, 4 place Jussieu,
          F 75252 PARIS CEDEX 05 (France).}
}
\centerline{\em Universit\'es Pierre et Marie Curie (Paris 6) et Denis
Diderot (Paris 7);} \centerline{\em Unit\'e de recherche associ\'ee au
CNRS (D0 280).}
\vskip 1.5cm

{\bf Abstract:}
I build $N=1$ superstrings in $\Bbb R^4$ out of purely geometric bosonic
data. The world-sheet is a bilayer of uniform thickness and the $2D$
supercharge vanishes in a natural way.
\smallskip

\vfill
\end{titlepage}
%
% texte du papier
%
\section{Introduction}
The usual approach to superstrings \cite{GSW} uses anticommuting
variables which are not very intuitive objects. In order to understand
them better, I have sought for a more pictorial description.
The basic idea is to use standard bosonization techniques \cite{AGBMNV}
and to interpret geometrically the compactified bosonic field
as kinks in the normal bundle. This is only possible when the
space-time is a four-manifold. The resulting model is the
following: I consider a bilayer with a uniform thickness living in a
four dimensional, flat Euclidean space
and choose an action proportional to the total area $A$ of this bilayer.
I show that this is a $\sigma$-model, taking values in the
projectified normal bundle, which can be fermionized into a
worldsheet Dirac fermion coupled to the normal connection \cite{Sp}.
For a particular value of the thickness, related to the string tension,
this model is equivalent to a free four-vector Majorana fermion with the
orthogonality constraint of a spinning string (the massless Dirac-Ramond
equation) \cite{R}.

\section{Action}

Our bilayers are described by:

\ $\star$ \ a smooth closed orientable $2D$ surface $\Sigma$, with $p$
marked points $S_1 \cdots S_p$\ ;

\ $\star$ \ an immersion $X:\Sigma \to {\Bbb R}^4$\ ;

\ $\star$ \ a smooth section of the projectified normal bundle induced
by $X$ on $\Sigma$ \\ ($Y\in\Gamma(PN_X \Sigma)$
can be singular at the punctures $S_1 \cdots S_p$)\ ;

\ $\star$ \ a thickness $2\delta>0$\ . \hfill\break

The $S_i$'s are the limits of infinitesimal circles mapped to
twisted strings. If $y(P)$ is a unit vector in the line
$Y(P) \ (\forall P\in\Sigma)$, the area of the bilayer
$(X\pm\delta y)(\Sigma)$ is:
	\begin{equation}
	\label{A1}
	A=\int_{\Sigma}d\xi^1\wedge d\xi^2\left\{\big(\mathrm{det}
	[\partial_a (X+\delta y).\partial_b(X+\delta y)]\big)^{1/2}
	+\big(\mathrm{det}[\partial_a(X-\delta y).\partial_b
	(X-\delta y)]\big)^{1/2}\right\}
	\end{equation}
which I expand in powers of $\delta$:
	\begin{equation}
	\label{A2}
	A=2\int_{\Sigma}d\xi^1\wedge d\xi^2 \ g^{1/2}\left(
	1+{\delta^2 \over 2}\
	g^{ab}\ \partial_a y^\perp .\partial_b y^\perp +\delta^2\ {\cal R}
	+{\cal O}(\delta^4)\right).
	\end{equation}
Here, $\xi=(\xi^1;\xi^2)$ is a local coordinate system on $\Sigma$, the dot
denotes the standard inner product in ${\Bbb R}^4$, $\partial_a y^\perp$ is
the normal part of $\partial_a y,\ g_{ab}=\partial_a X.\partial_b X,\
g=\mathrm{det}[g_{ab}]$, and ${\cal R}$ is Ricci's scalar curvature. The
${\cal O}(\delta^4)$ terms, containing more derivatives, are irrelevant,
and I drop the topological term $\int_{\Sigma}d\xi^1\wedge d\xi^2 \ g^{1/2}
\ {\cal R}=8\pi(1-$genus$(\Sigma)).$
The second term in (\ref{A2}) can be rewritten as follows.
Pick a generic $N\in\Gamma(N_X\Sigma)$ with isolated zeros $Z_1 \cdots Z_q$
of indices $\iota_1 \cdots \iota_q$. The normal $n={N / \Vert N \Vert}$ and
binormal $b$ define a right handed orthonormal frame in $N_X\Sigma$ over
$\Sigma_Z=\Sigma\setminus\{Z_1 \cdots Z_q\}$, where
the normal connection $\nabla^{\perp}$ is represented by the
matrix $\pmatrix{d&-T\cr T&d\cr}$ with \
$d=d\xi^1 \partial_1+ d\xi^2 \partial_2$\ \ and \ \ $T=b.dn$\ .
If $\theta :\Sigma_Z \to {\Bbb R}/\pi {\Bbb Z}$ is the angle from
$\pm n$ to $Y$, we have:
	\begin{eqnarray}
	\pm y &=& \cos\theta \ n+\sin\theta \ b \ ,\nonumber \\
	dy^\perp &=& \pm(d\theta+T)\ (\cos\theta \ b-\sin\theta \ n)\ ,
	\nonumber \\
	A &=& 2\int_{\Sigma}d\xi^1\wedge d\xi^2 \ g^{1/2}+\delta^2\int_\Sigma
	\omega\wedge *\omega \ ,
	\end{eqnarray}
where $\omega=*(d\theta+T)\ (=(\partial_1\theta+T_1)d\xi^2-
(\partial_2\theta+T_2)d\xi^1$\ \ if \ \ $g_{ab}=e^\phi \delta_{ab}$)\ .\\
I take the action to be $S=\mu A$\ , $\mu$ being the string tension of
one layer. In the partition function
$\cal Z (X)=\int D\theta \ e^{-\mu \delta^2
\int_\Sigma \omega\wedge *\omega}$,
we sum over the $\theta$'s which satisfy $\oint_{Z_j}\omega=0$, since
$Y$ is regular at these points, and $\oint_{S_i}\omega=n_i \pi \
(n_i \in {\Bbb Z})$ (the boundary strings can be twisted).
Among these functions, the classical configurations are the solutions
of the equation of motion $d\omega=0$
and are parametrized by $H_1(\Sigma;{\Bbb Z})$.

\section{Fermions}
Since $PN_X\Sigma$ is a circle bundle, this system admits kinks and a
fermionic representation by holonomies \cite{Sk}.
If $\gamma:[0;1]\to\Sigma$ is a path joining $P_0$ to $P$, we define:
	\begin{eqnarray}
	\label{bc}
	b &=& exp\ (k\int_\gamma id\theta-\omega)\qquad
	c  = exp\ (-k\int_\gamma id\theta-\omega)\\
	\bar b &=& exp\ (k\int_\gamma id\theta+\omega)\nonumber\qquad
	\bar c = exp\ (-k\int_\gamma id\theta+\omega)\ .
	\end{eqnarray}
Due to the equation of motion ($d\omega=0$)\ , their correlators only depend
on $[\gamma]\in H_1(\Sigma,P-P_0;{\Bbb Z})$.
In order to recover
	\begin{equation}
        \label{propagator}
	{1\over \cal Z (X)}\int D\theta \
	e^{-\mu \delta^2 \int_\Sigma\omega\wedge *\omega}\ b(z) c(0)=
	\langle b(z) c(0)\rangle \sim z^{-1}\ ,
	\end{equation}
on $\Bbb C$ and without the gauge field $T$, we must fix
$k=(2\pi\mu\delta^2)^{1/2}$, as can be seen after a Gaussian
integration. Moreover, for the special value $k=1$\ , i.e.
$\delta=(2\pi\mu)^{-1/2}=\delta_0$\ ,
there is no quartic term in the fermionic action \cite{CG} and
$\psi=\pmatrix{c \cr \bar b \cr}$ satisfies the following
equation of motion:
	\begin{equation}
	\label{Dirac}
	\pmatrix{0& 2\partial+i(T_1 +iT_2)\cr
	2\bar\partial+ i(T_1 -iT_2) &0} \pmatrix{c \cr \bar b \cr}=
	(\partial \!\!\!/ +i T\!\!\!\!/)\psi=0 \ .
	\end{equation}
This shows that $\psi$ is a $2D$ Dirac spinor and a vector in $N_X\Sigma$\ :
	\begin{equation}
	\label{psi}
	\psi\in\Gamma (K^{1/2}\otimes_{\Bbb C} N_X\Sigma) \oplus
	\Gamma (K^{-1/2}\otimes_{\Bbb C} N_X\Sigma)\ .
	\end{equation}
Here, $N_X\Sigma$ is viewed as a complex line bundle on $\Sigma$\ ,
$K$ denotes the canonical line bundle of holomorphic $(1,0)$-forms on
$\Sigma$\ , $K^{1/2}$ is one of the $2^{2\mathrm{genus}(\Sigma)+p}$ spin
structures on $\Sigma$ \ \cite{AGBMNV}\ , $K^*$ is the
dual bundle of $K$ and $K^{-1/2}=K^{1/2}\otimes_{\Bbb C} K^*$\ .
Since the normal connection $\nabla^\perp$ is the projection on $N_X\Sigma$
of the trivial connection $\nabla$ acting on sections of the total bundle
$X^*(T{\Bbb R}^4)=T\Sigma\oplus^\perp_{\Bbb R} N_X\Sigma$,
we can replace $\psi$ by a free four-vector Majorana fermion
	\begin{equation}
	\label{FDF}
	\Psi\in\Gamma (K^{1/2}\otimes_{\Bbb R} X^*(T{\Bbb R}^4))\oplus
	\Gamma (K^{-1/2}\otimes_{\Bbb R} X^*(T{\Bbb R}^4))
	\ \ \mathrm{and} \ \
	\partial \!\!\!/ \Psi=0 \ ,
	\end{equation}
with the orthogonality constraint $\Psi.dX=0$ to be applied on the
Hilbert space in order to recover the same number of degrees of freedom
in (\ref{psi}) and (\ref{FDF}).
We thus obtain three equivalent descriptions of a fermionic string
satisfying the (massless) Dirac-Ramond equation:

\ $\star$ \ a $\sigma$-model in $PN_X \Sigma$ \ ;

\ $\star$ \ $\psi\in\Gamma (K^{1/2}\otimes_{\Bbb C} N_X\Sigma)\oplus
\Gamma (K^{1/2}\otimes_{\Bbb C} N_X\Sigma) \ \ $and$
\ \ (\partial \!\!\!/ +i T\!\!\!\!/)\psi=0$ \ ;

\ $\star$ \ $\Psi\in\Gamma (K^{1/2}\otimes_{\Bbb R} X^*(T{\Bbb R}^4))
\oplus \Gamma (K^{-1/2}\otimes_{\Bbb R} X^*(T{\Bbb R}^4)),\
\Psi \ $is real$ \ , \
\partial \!\!\!/ \Psi=0\ $ and $\ \Psi.dX=0 $\ .

\section{Conclusion}

The previous computations suggest a simple picture for superstrings
in four dimensions: they are double covers of bosonic strings and the
two nearby world-sheets must be separated by $2\delta_0$ in order to
have free fields. This suggests that one interpret the tachyonic
instability of bosonic strings as a phase transition to a fermionic vacuum.

\vskip 2cm
\begin{em}
\underline {Acknowledgements}:
I thank Jean-Beno\^\i t Bost and Krzysztof Gawedzki for helpful discussions.
\end{em}

%\vfill\eject


\begin{thebibliography}{20}

\bibitem{GSW}
M.B. Green, J.H. Schwarz and E. Witten: ``Superstring Theory''.\hfill\break
(Cambridge University Press, 1987).

\bibitem{AGBMNV}
L. Alvarez-Gaum\'e, J.-B. Bost, G. Moore, P. Nelson and C. Vafa: \hfill\break
``Bosonization on higher genus Riemann surfaces''.\hfill\break
Commun. Math. Phys. {\bf 112}, 503-552 (1987).

\bibitem{Sp}
M.D. Spivak: ``A comprehensive introduction to differential geometry''.
\hfill\break (Publish or Perish, 1979).

\bibitem{R}
P. Ramond: ``Dual theory for free fermions''.
Phys. Rev. {\bf D3}, 2415-2418 (1971).

\bibitem{Sk}
T.H.R. Skyrme: ``Kinks and the Dirac equation''.
J. Math. Phys. {\bf 12} (8), 1735-1743 (1971).

\bibitem{CG}
E.Charpentier and K.Gawedzki:
``Bosonization in background of metric and gauge''.\hfill\break
J. Math. Phys. {\bf 34} (2), 381-436 (1993).


\end{thebibliography}
\end{document}